\documentclass[aps,pra,twocolumn,superscriptaddress]{revtex4}

\usepackage{color}
\usepackage{amsmath,amsfonts,amssymb}
\usepackage{graphicx}
\usepackage{algorithm}
\usepackage{algpseudocode}
\usepackage{xr}
\usepackage{xcite}
\usepackage{float}
\usepackage{bm,bbm}
\usepackage{mathrsfs} 

\usepackage{epsfig}
\usepackage{verbatim}
\usepackage{array}
\usepackage{epstopdf}
\usepackage{dcolumn}
\usepackage{tabularx}
\usepackage[unicode=true,bookmarks=true,bookmarksnumbered=false,bookmarksopen=false,breaklinks=false,pdfborder={0 0 1},backref=false,colorlinks=true]{hyperref}
\hypersetup{linkcolor=magenta, urlcolor=blue, citecolor=blue, pdfstartview={FitH}, hyperfootnotes=false, unicode=true}

\usepackage{tikz}

\newcommand*{\black}{\textcolor{black}} 

\usepackage[T1]{fontenc}
\usepackage{tgtermes}

\begin{document}

\title{Generalizable control for quantum parameter estimation through reinforcement learning}
\author{Han Xu}
\affiliation{Department of Physics, City University of Hong Kong, Tat Chee Avenue, Kowloon, Hong Kong SAR, China, and City University of Hong Kong Shenzhen Research Institute, Shenzhen, Guangdong 518057, China}
\affiliation{School of Physics and Technology, Wuhan University, Wuhan 430072, China}
\author{Junning Li}
\affiliation{Department of Physics, City University of Hong Kong, Tat Chee Avenue, Kowloon, Hong Kong SAR, China, and City University of Hong Kong Shenzhen Research Institute, Shenzhen, Guangdong 518057, China}
\author{Liqiang Liu}
\affiliation{Department of Mechanical and Automation Engineering, The Chinese University of Hong Kong, Shatin, Hong Kong SAR, China}
\author{Yu Wang}
\affiliation{School of Physics and Technology, Wuhan University, Wuhan 430072, China}
\author{Haidong Yuan}
\email{hdyuan@mae.cuhk.edu.hk}
\affiliation{Department of Mechanical and Automation Engineering, The Chinese University of Hong Kong, Shatin, Hong Kong SAR, China}
\author{Xin Wang}
\email{x.wang@cityu.edu.hk}
\affiliation{Department of Physics, City University of Hong Kong, Tat Chee Avenue, Kowloon, Hong Kong SAR, China, and City University of Hong Kong Shenzhen Research Institute, Shenzhen, Guangdong 518057, China}

\begin{abstract}
  Measurement and estimation of parameters are essential for science and engineering, where one of the main quests is to find systematic schemes that can achieve high precision. While conventional schemes for quantum parameter estimation focus on the optimization of the probe states and measurements, it has been recently realized that control during the evolution can significantly improve the precision. The identification of optimal controls, however, is often computationally demanding, as typically the optimal controls depend on the value of the parameter which then needs to be re-calculated after the update of the estimation in each iteration. Here we show that reinforcement learning provides an efficient way to identify the controls that can be employed to improve the precision. We also demonstrate that reinforcement learning is highly \black{generalizable}, namely the neural network trained under one particular value of the parameter can work for different values within a broad range. These desired features make reinforcement learning \black{an efficient alternative to} conventional optimal quantum control methods.
\end{abstract}

\maketitle

\section{Introduction}
Metrology, which studies high precision measurement and estimation, has been one of the main driving forces in science and technology. Recently, quantum metrology, which uses quantum mechanical effects to improve the precision, has gained increasing attention for its potential applications in imaging and spectroscopy  \cite{kolobov1999spatial,lugiato2002quantum,morris2015imaging,roga2016security,tsang2016quantum,Giovannetti2011}.

One of the main quests in quantum metrology is to identify the highest precision that can be achieved with given resources. \black{Typically the desired parameter, $\omega$, is encoded in a dynamics ${\it \Lambda}_{\omega}$. After an initial probe state $\rho_0$ is prepared, the parameter is encoded in the output state as $\rho_\omega={\it \Lambda}_{\omega}(\rho_0)$. Proper measurements on the output state then reveals the value of the parameter}. To achieve the highest precision, one needs to optimize the probe states, the controls during the dynamics and the measurements on the output states. Previous studies have been mostly focused on the optimization of the probe states and measurements  \cite{Giovannetti2011}. The control only starts to gain attention recently  \cite{yuan2015optimal, yuan2016sequential,Pang2017a,Pang2017b,Liu2017,Liu2017a,Yang2017,Naghiloo2017,Sekatski2017,Braun2017, degen2017, braun2018}. It has now been realized that properly designed controls can significantly improve the precision limits. The identification of optimal controls, however, is often highly complicated and time-consuming. This issue is particularly severe in quantum parameter estimation, as typically optimal controls depend on the value of the parameter, which can only be estimated from the measurement data. When more data are collected, the optimal controls also need to be updated, which is conventionally achieved by another run of the optimization
algorithm
. This creates a high demand for the identification of efficient algorithms to find the optimal controls in quantum parameter estimation.

Over the past few years, machine learning has demonstrated astonishing achievements in certain
high-dimensional input-output problems, such as playing video games  \cite{Mnih2015} and mastering the game of Go \cite{Silver2016}. 
\black{Reinforcement Learning (RL) \cite{sutton2018} is one of the most basic yet powerful paradigms of machine learning. In RL, an agent interacts with an environment with certain rules and goals set forth by the problem desired. By trial and error, the agent optimizes its strategy to achieve the goals, which is then translated to a solution to the problem. RL has been shown to provide improved solutions to many problems related to quantum information science, including quantum state transfer \cite{Zhang2018},  quantum error correction \cite{Fosel2018}, quantum communication \cite{Wallnofer2019},  quantum control \cite{Bukov2018,Niu2019,An2019} and experiment design  \cite{Melnikov2018}}.

Here we show that RL \black{serves as an efficient alternative to identify controls that are helpful} in quantum parameter estimation. A main advantage of RL is that it is highly generalizable, i.e., the agent trained through RL under one value of the parameter works for a broad range of the values. There is then no need for re-training after the update of the estimated value of the parameter from the accumulated measurement data, \black{which makes the procedure less resource-consuming under certain situations}.

\section{Results}


We consider a generic control problem described by the Hamiltonian  \cite{Khaneja2005}:
\begin{equation}
\hat{H}(t)=\hat{H}_0(\omega)+\sum_{k=1}^{p}u_k(t)\hat{H}_k,
\end{equation}
where $\hat{H}_0$ is the time-independent free evolution of the quantum state,
$\omega$ the parameter to be estimated, $u_k(t)$ the $k$th time-dependent control field,
$p$ the dimensionality of the control field,
and $\hat{H}_k$ couples the control field to the state.

The density operator of a quantum state (pure or mixed) evolves according to \black{the master equation} \cite{breuer2002theory},
\begin{equation}
    \partial_t\hat{\rho}(t)=-i\left[\hat{H}(t),\hat{\rho}(t)\right]+{\it \Gamma}\left[\hat{\rho}(t)\right],
    \label{eq:meq}
\end{equation}
where ${\it \Gamma}[\hat{\rho}(t)]$ indicates a noisy process, the detailed form of which depends on
the specific noise mechanism and will be detailed later.

The key quantity in quantum parameter estimation is the QFI  \cite{helstrom1976quantum,Holevo,Petz2010,Braunstein1994}, defined by
\begin{equation}
F(t)=\mathrm{Tr}\left[\hat{\rho}(t)\hat{L}_s^2(t)\right],
\label{eq:qfi}
\end{equation}
where $\hat{L}_s(t)$ is the so-called symmetric logarithmic derivative that can be obtained by solving the
equation $\partial_{\omega}\hat{\rho}(t)=\frac{1}{2}\left[\hat{\rho}(t)\hat{L}_s(t)+\hat{L}_s(t)\hat{\rho}(t)\right]$
 \cite{helstrom1976quantum,Holevo,Braunstein1996}. According to the Cram\'{e}r-Rao bound, the QFI provides a saturable lower bound on the estimation as $\delta \hat{\omega}\geq \frac{1}{\sqrt{nF(t)}}$, where $\delta \hat{\omega}=\sqrt{E[(\hat{\omega}-\omega)^2]}$ is the standard deviation of an unbiased estimator $\hat{\omega}$, and $n$ is the number of times the procedure is repeated. Our goal is therefore to search for optimal control sequences $u_k(t)$ that maximize the QFI at time $t=T$ (typically the conclusion of the control), $F(T)$,
respecting all constraints possibly imposed in specific problems.
Practically, we consider piecewise constant controls so the total evolution time $T$ is discretized into
$N$ steps with equal length $\Delta T$ labeled by $j$, and we use $u_k^{(j)}$ to denote the strength of the control field $u_k$ on the $j$th time step.
Researches of such problem are frequently tackled by the Gradient Ascent Pulse Engineering (GRAPE) method \cite{Khaneja2005},
which searches for an optimal set of control fields by updating their values according to the gradient of a cost function encapsulating the goal of the optimal control. It has been found that GRAPE is successful in preparing optimal control pulse sequences that improve the precision limit of quantum parameter estimation in noisy processes \cite{Liu2017,Liu2017a}. 
Many alternative algorithms can tackle this optimization problem such as the stochastic gradient ascent(descent) method and microbial genetic algorithm \cite{Harvey2011},
but the convergence to the optimal control fields becomes much slower when the dimensionality ($p$) of the control field or the discretization steps ($N$) increases. Other optimal quantum control algorithms, such as Krotov's method \cite{Sklarz2002,Palao2003,Machnes2011,Reich2012,Goerz2015} and CRAB algorithm \cite{Doria2011}, typically depend on the value of the parameter, thus need to be run repeatedly along the update of the estimation, which is highly time-consuming. More efficient algorithms are thus highly desired.

\begin{figure}
  \centering
  \includegraphics[width=0.9\linewidth]{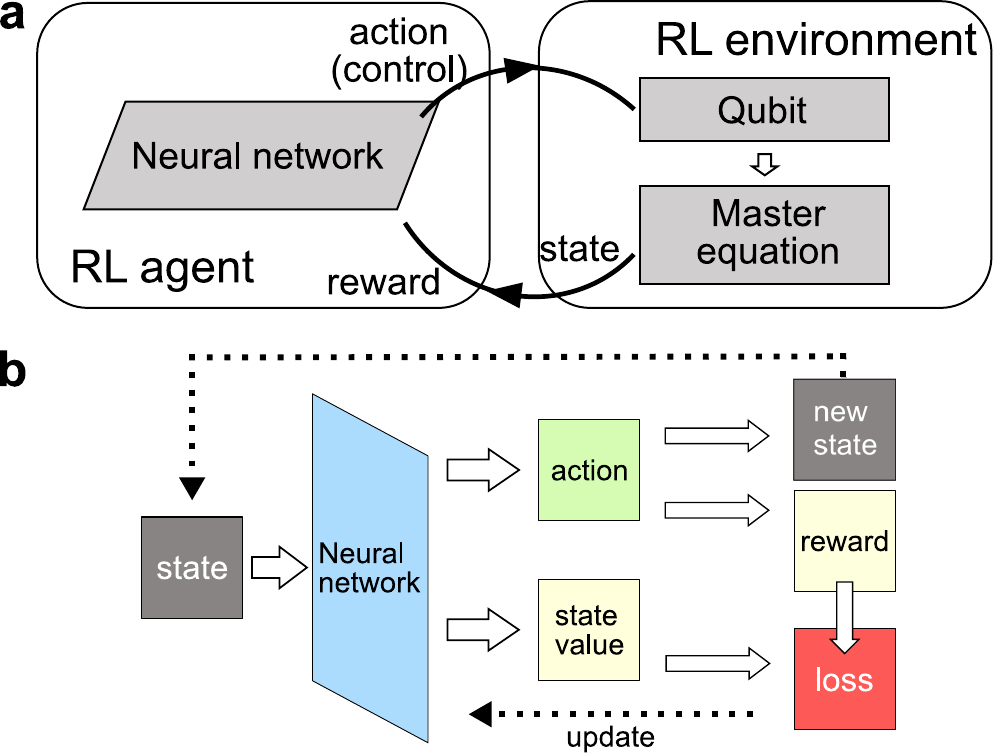}
  \caption{ \text{Schematics of the reinforcement learning procedure.} \textbf{a} the RL agent-environment interaction as a Markovian decision process. The RL agent who first takes an action is prescribed by a neural network. The action is essentially the control field which steers the qubit. Then, depending on the consequence of the action, the agent receives a reward. \black{\textbf{b} Schematic flow chart of one training step of the Actor-Critic algorithm. The hollow arrows show the data flow of the algorithm, and the dotted arrows show updates of the states and the neural network. In each time step, the state evolves according to the action chosen by the neural network, generating a new state which is used as the input to the network in the next time step. The loss function (detailed in Methods and the Supplementary Methods) is used to update the parameters of the neural network so as to optimize its choice of actions.  The procedure is repeated until actions in all time steps are generated, forming the full evolution of the state and concluding one training episode.}
  \label{fig:RL}
  }
\end{figure}

\begin{figure*}
  \centering
  \includegraphics[width=0.75\linewidth]{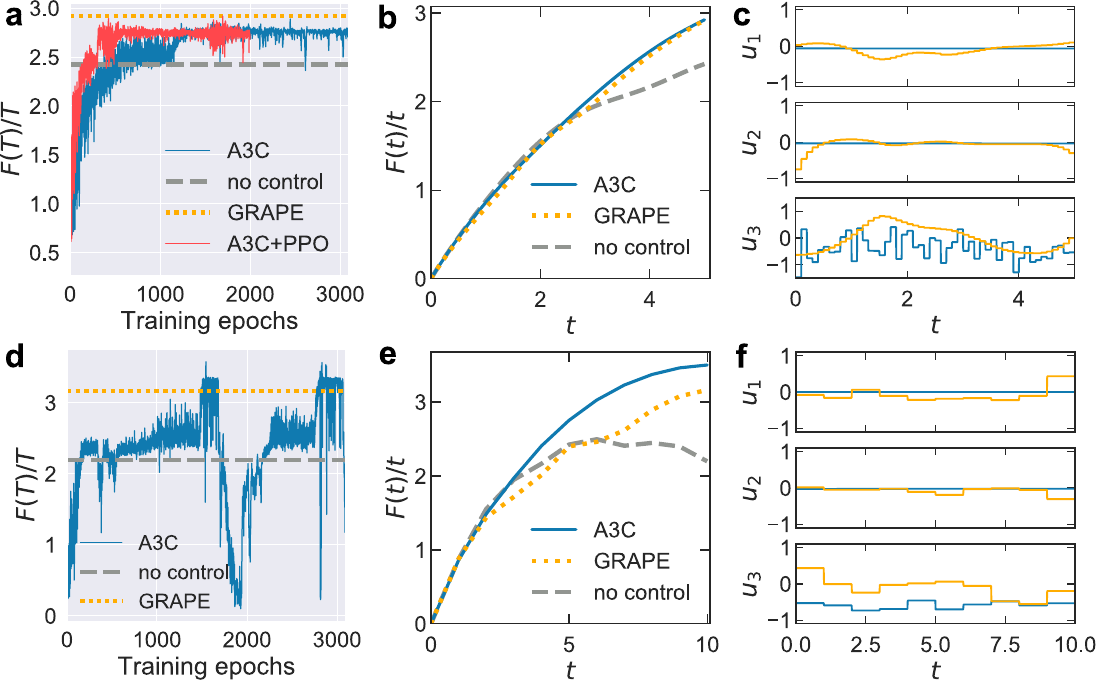}
  \caption{ \text{Quantum parameter estimation under dephasing dynamics with $\bm{\vartheta={\mathrm \pi}/4}$, $\bm{\phi=0}$ using square pulses.} \textbf{a}-\textbf{c} results for $\Delta T=0.1$, $T=5$. \textbf{d}-\textbf{f} results for $\Delta T=1$, $T=10$. \textbf{a}, \textbf{d} show the learning procedure, namely $F(T)/T$ as functions of training epochs. \textbf{b}, \textbf{e} show $F(t)/t$  for one of the best training results selected from \textbf{a} and \textbf{d} respectively. \textbf{c} and \textbf{f} show the pulse profiles corresponding to \textbf{b} and \textbf{e}.
  \label{fig:main:dp}
  }
\end{figure*}

In this work, we employ RL to solve the problem \black{and compare the results to GRAPE. Our implementation of GRAPE follows Ref.~\cite{Liu2017}}.
Figure~\ref{fig:RL} shows schematics of the RL procedure and the Actor-Critic algorithm \cite{sutton2018} used in this work. In order to improve the efficiency of computation, we used a parallel version of the Actor-Critic algorithm called Asynchronous Advantage Actor-Critic (A3C) algorithm \cite{Mnih2016}. For more extensive reviews of RL, Actor-Critic algorithm and A3C, see Methods and the Supplementary Methods.

Next we apply the algorithm to two commonly considered noisy processes: dephasing and spontaneous emission, to demonstrate the effect of the algorithm.

\subsection{Dephasing Dynamics}

Under dephasing dynamics, the master equation, Eq.~\eqref{eq:meq}, takes the following form \cite{Liu2017}:
\begin{equation}
    \partial_t\hat{\rho}(t)=-i\left[\hat{H}(t),\hat{\rho}(t)\right]
    +\frac{\gamma}{2}\left[\hat{\sigma}_{\mathbf n}\hat{\rho}(t)\hat{\sigma}_{\mathbf n}-\hat{\rho}(t)\right],
    \label{eq:dephase}
\end{equation}
where
\begin{equation}
\hat{H}(t)=\frac{1}{2}\omega_0\hat{\sigma}_3+{\mathbf u}(t)\cdot{\bm \sigma},
    \label{eq:genham}
\end{equation}
the control field $\mathbf{u}(t)=(u_1,u_2,u_3)$ is a magnetic field that couples to
${\bm \sigma}=(\hat\sigma_1,\hat\sigma_2,\hat\sigma_3)$, and $\gamma$ is the dephasing rate which is taken as 0.1 throughout the paper.
We consider a dephasing along a general direction given by
${\mathbf n}=(\sin{\vartheta}\cos{\phi},\sin{\vartheta}\sin{\phi},\cos{\vartheta})$,
$\hat\sigma_{\mathbf n}={\mathbf n}\cdot{\bm \sigma}$.
The parameter to be estimated is $\omega_0$ in Eq.~\eqref{eq:genham},
the true value of which is assumed to be 1, and we take $\omega_0^{-1}=1$ as our time unit.
We choose the probe state, i.e. the initial state of the evolution,
as $(|0\rangle+|1\rangle)/\sqrt{2}$ in all subsequent calculations, where
$|0\rangle,|1\rangle$ are the eigenstates of $\hat\sigma_3$.

\begin{figure}
  \centering
  \includegraphics[width=0.9\linewidth]{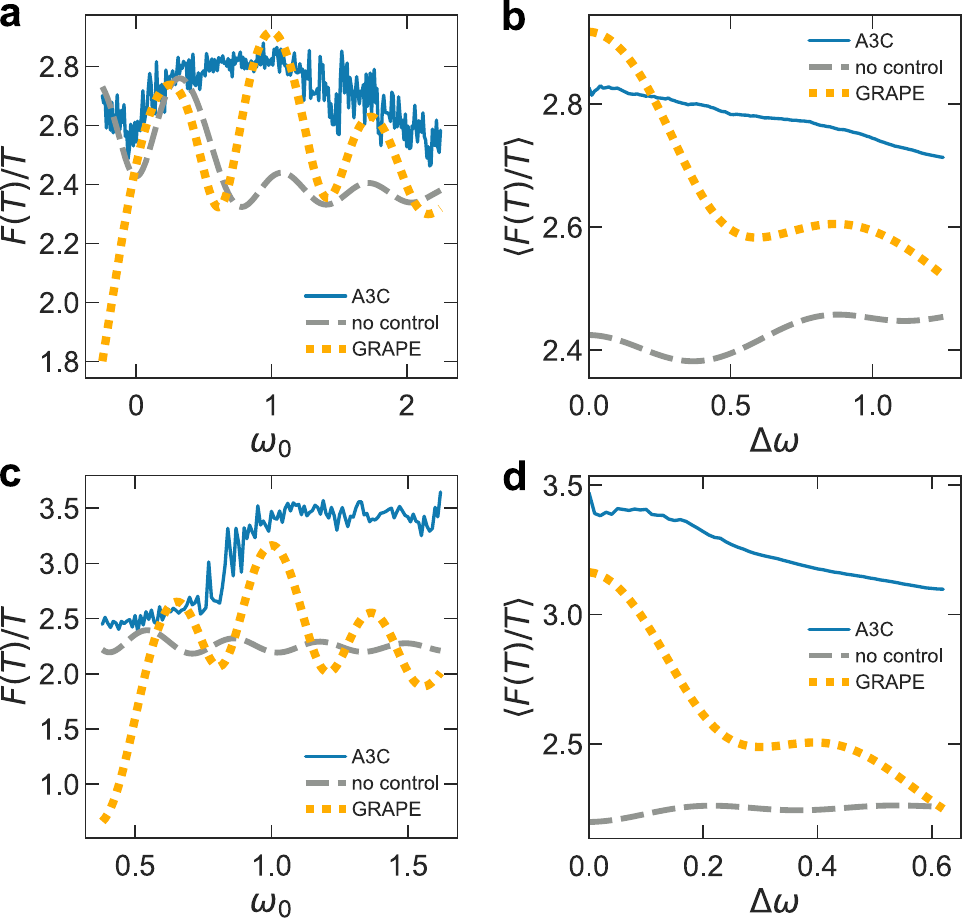}
  \caption{ \text{\black{Generalizability} of the control under dephasing dynamics.} \textbf{a}, \textbf{c} $F(T)/T$ v.s. $\omega_0$ for three different methods. Note that the results from the GRAPE method are obtained using the pulses generated for $\omega_0=1$ only, while those from A3C are obtained using a neural network trained at $\omega_0=1$. \textbf{b}, \textbf{d} average $F(T)/T$ in a range $[1-\Delta\omega,1+\Delta\omega]$ corresponding to the results of \textbf{a} and \textbf{c} respectively. \textbf{a}, \textbf{b} $\Delta T=0.1$, $T=5$; \textbf{c}, \textbf{d} $\Delta T=1$, $T=10$.
  \label{fig:main:xfer_dp}
  }
\end{figure}

In Fig.~\ref{fig:main:dp} we present our numerical results on QFI under dephasing dynamics with $\vartheta={\mathrm \pi}/4$, $\phi=0$ using square pulses. Figure~\ref{fig:main:dp}{a}-{c} show the results for $\Delta T=0.1$.
Figure~\ref{fig:main:dp}{a} shows the training process in terms of $F(T)/T$ as functions of the number of training epochs. The blue line shows results from the training using A3C algorithm. The value of $F(T)/T$ corresponding to results from GRAPE and the case with no control are shown as the orange dotted line and grey dashed line, respectively. The red line shows results from ``A3C+PPO'', an enhanced version of A3C which converges faster \cite{Schulman2017}. The details of this algorithm is explained in the Supplementary Methods. We can see that after sufficient training epochs, results from A3C exceed that for the case with no control, and approaches the optimal results found by GRAPE. On the other hand, ``A3C+PPO'' converges more quickly to essentially the same result of A3C.

We select one training outcome from those with best performances in Fig.~\ref{fig:main:dp}{a} and show $F(t)/t$ and the pulse profiles in
Fig.~\ref{fig:main:dp}{b}, {c} respectively. As can be seen from Fig.~\ref{fig:main:dp}{b}, both GRAPE and A3C outperform the case with no control, while the results of A3C are comparable to those from GRAPE.

Figure~\ref{fig:main:dp}{d}-{f} show results with a larger time step, $\Delta T=1$. From the training results shown in  Fig.~\ref{fig:main:dp}{d}, we see that results from A3C occasionally exceed those from GRAPE, for example at training epoch approximately 1600 and 3000. $F(t)/t$ and the pulse profile of one of the best performing results is again shown in Fig.~\ref{fig:main:dp}{e} and {f}, and we see from Fig.~\ref{fig:main:dp}{e} that A3C indeed outperforms GRAPE in this case.

We have discussed dephasing dynamics along a particular axis pertaining to Fig.~\ref{fig:main:dp}, and the results for several other dephasing axes are shown in the Supplementary Discussion. We conclude from these results that in most cases, the A3C algorithm is capable to produce results comparable to those from GRAPE, while in selected situations (e.g. larger $\Delta T$) A3C may outperform GRAPE.

We now discuss the \black{generalizability} of the control sequences for quantum parameter estimation, a key result of this paper. Since the true value of $\omega_0$ is not known \emph{a priori}, the control sequence has to be found optimal for a chosen $\omega_0$. When such sequence is applied in situations under other $\omega_0$ values, the true value is still measured, but the resulting QFI is lower than when the optimal control for true $\omega_0$ is used. \black{In order to raise the QFI, one must then perform a second measurement using control sequences optimized for the estimated true value of $\omega_0$. The entire procedure therefore involves two steps, using different pulse sequences. This is fundamentally different than other typical measurements in quantum control, e.g. evaluation of fidelities of quantum gates \cite{Goerz2014}, for which there is no need for a second pulse sequence or a second measurement}.

The dotted lines in the left column of Fig.~\ref{fig:main:xfer_dp} show the QFI resulting from measurements with the optimal control found for $\omega_0=1$ with GRAPE. Results without control are shown as grey dashed lines for comparison. The range of $\omega_0$ covers a period of $2{\mathrm \pi}/T$. As expected, the QFI is largest at $\omega_0=1$, but reduces as $\omega_0$ deviates from 1. As $\omega_0$ further varies, the QFI increases at some values of $\omega_0$ which may be due to the geometric relationship of the phase that corresponding to those $\omega_0$ values and the phase at $\omega_0=1$. In any case, these QFI values are consistently lower than  the value at $\omega_0=1$. An obvious way to improve the QFI is to generate new optimal control sequences for each value of $\omega_0$ from GRAPE, but this is costly as the computational complexity scales as ${\cal O}(N^3)$. A detailed discussion on the computational complexity can be found in Supplementary Discussion. 

With A3C we have an efficient solution to this problem. We can train the neural network at $\omega_0=1$, and use this particular network to generate control sequences for different $\omega_0$ values. The neural network is only trained at $\omega_0=1$. However, the trained neural network works for a broad range of parameter values. There is no need to re-train the neural network with the updated estimation of the parameter. The computational cost is thus simply ${\cal O}(N)$ so it is much more efficient than generating new sequences with GRAPE. These results from A3C are shown in the left column of Fig.~\ref{fig:main:xfer_dp} as blue solid lines which represents the best-performing sequence from 100 trials generated from the trained neural network. For $\Delta T=0.1$ (Fig.~\ref{fig:main:xfer_dp}{a}), although the QFI in the training 
$\omega_0=1$ 
is slightly lower for A3C than that of GRAPE, A3C demonstrates higher \black{generalizability} as the QFI deceases slowly when $\omega_0$ deviates from 1.
For $\Delta T=1$ (Fig.~\ref{fig:main:xfer_dp}{c}), the QFI of A3C is consistently higher than GRAPE except a narrow range of $\omega_0$ around 0.65.

To further reveal the \black{generalizability} of different methods, we consider the measurement in an ensemble with $\omega_0$ uniformly distributed in $[1-\Delta\omega,1+\Delta\omega]$. The performance of the quantum parameter estimation is therefore given by the average $F(T)/T$,
\begin{equation}
    \langle F(T)/T\rangle=\frac{1}{2\Delta\omega}\int_{1-\Delta\omega}^{1+\Delta\omega}F(T)/T~d\omega.
    \label{eq:ratio}
\end{equation}
These results are shown in the right column of Fig.~\ref{fig:main:xfer_dp}, which are averages of the data in the corresponding panels in the left column. As seen from Fig.~\ref{fig:main:xfer_dp}{b} ($\Delta T=0.1$), $\langle F(T)/T\rangle$ for GRAPE is high at small $\Delta\omega$ but drops quickly as $\Delta\omega$ is increased. On the contrary, $\langle F(T)/T\rangle$ for A3C is lower than that for GRAPE at small $\Delta\omega$, but decays much more slowly. As a consequence, $\langle F(T)/T\rangle$ for A3C exceeds that for GRAPE beyond $\Delta\omega\gtrsim0.22$. This result indicates that for measurements involving a reasonably varying parameter, A3C demonstrates higher \black{generalizability}. For $\Delta T=1$, the results of A3C always exceed GRAPE as seen from Fig.~\ref{fig:main:xfer_dp}{d}. The result for A3C decays much more slowly than that for GRAPE, in consistency with the $\Delta T=0.1$ case. 

Intuitively without control and noise, the optimal strategy is preparing the initial probe state as $(|0\rangle+|1\rangle)/{\sqrt{2}}$, since this state has the fastest rate of rotations under the Hamiltonian. Since the evolution of the state is also affected by dephasing, competitions exist between the parametrization and the effect of noise. When the evolution time is short, the parametrization dominates, in which case the control does not help much. However, in experimentally relevant situations the evolution time is typically long enough for noises to dominate. The controls are therefore useful as they can steer the states to regions where those states are less affected by the noise, even if such states may have a slower speed of parametrization. GRAPE and RL-based methods are both systematical ways to find controls, however, as we have demonstrated, A3C is more \black{generalizable}.

\subsection{Spontaneous Emission}

\begin{figure*}
  \centering
  \includegraphics[width=0.75\linewidth]{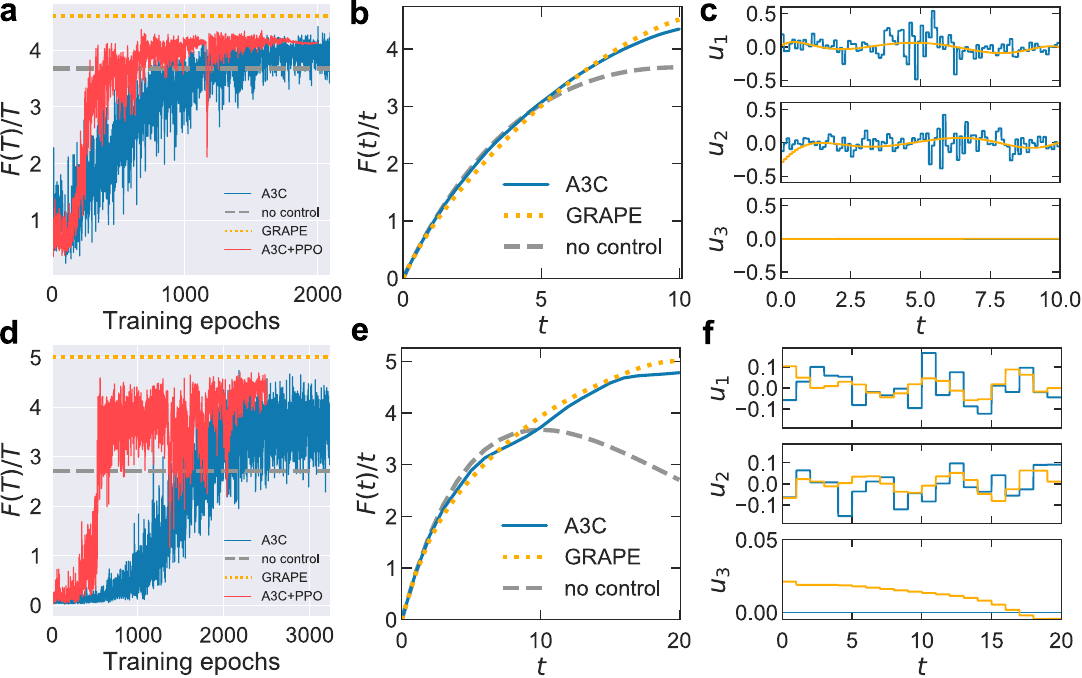}
  \caption{ \text{Quantum parameter estimation under spontaneous emission using square pulses.} \textbf{a}-\textbf{c} results for $\Delta T=0.1$, $T=10$. \textbf{d}-\textbf{f} results for $\Delta T=1$, $T=20$. \textbf{a}, \textbf{d} show the learning procedure. \textbf{b} and \textbf{e} show $F(t)/t$ for one of the best training results selected from \textbf{a} and \textbf{d} respectively. \textbf{c} and \textbf{f} show the pulse profiles corresponding to \textbf{b} and \textbf{e}.
  \label{fig:main:se}
  }
\end{figure*}

A process involving the spontaneous emission is
described by the Lindblad master equation \cite{Liu2017}:
\begin{equation}
    \begin{split}
    \partial_t\hat\rho(t)=&-i\left[\hat{H}(t),\hat\rho(t)\right] + \gamma_+\left[\hat\sigma_+\hat\rho(t)\hat\sigma_-
    -\frac{1}{2}\left\lbrace\hat\sigma_-\hat\sigma_+,\hat\rho(t)\right\rbrace\right]\\
        &+ \gamma_-\left[\hat\sigma_-\hat\rho(t)\hat\sigma_+ -\frac{1}{2}\left\lbrace\hat\sigma_+\hat\sigma_-,\hat\rho(t)\right\rbrace\right],
    \end{split}
    \label{eq:emission}
\end{equation}
where $\hat\sigma_{\pm}=(\hat\sigma_1\pm i\hat\sigma_2)/2$
and $\hat{H}$ is defined as Eq.~\eqref{eq:genham}. The \black{relaxation} rates are taken as $\gamma_+=0.1,\gamma_-=0$ throughout our discussion.

Figure~\ref{fig:main:se} shows numerical results on QFI with spontaneous emission. Figure~\ref{fig:main:se}{a}-{c} are for $\Delta T=0.1$, $T=10$, and Fig.~\ref{fig:main:se}{d}-{f} show calculations with a larger time step $\Delta T=1$, $T=20$. Figure~\ref{fig:main:se}{a}, {d} [left column] show the A3C training processes, in which the results from GRAPE are indicated as orange dotted line for reference. We see that ``A3C+PPO" converges faster, and both A3C and ``A3C+PPO" saturate to values slightly lower than GRAPE. Again, one of the best performing control is picked out  and the corresponding $F(t)/t$ and pulse profiles are shown in the middle and right column respectively. From Fig.~\ref{fig:main:se}{b}, {e} we see that for the best result from A3C, the QFI is lower than, but comparable to results from GRAPE.

\begin{figure}
  \centering
  \includegraphics[width=0.9\linewidth]{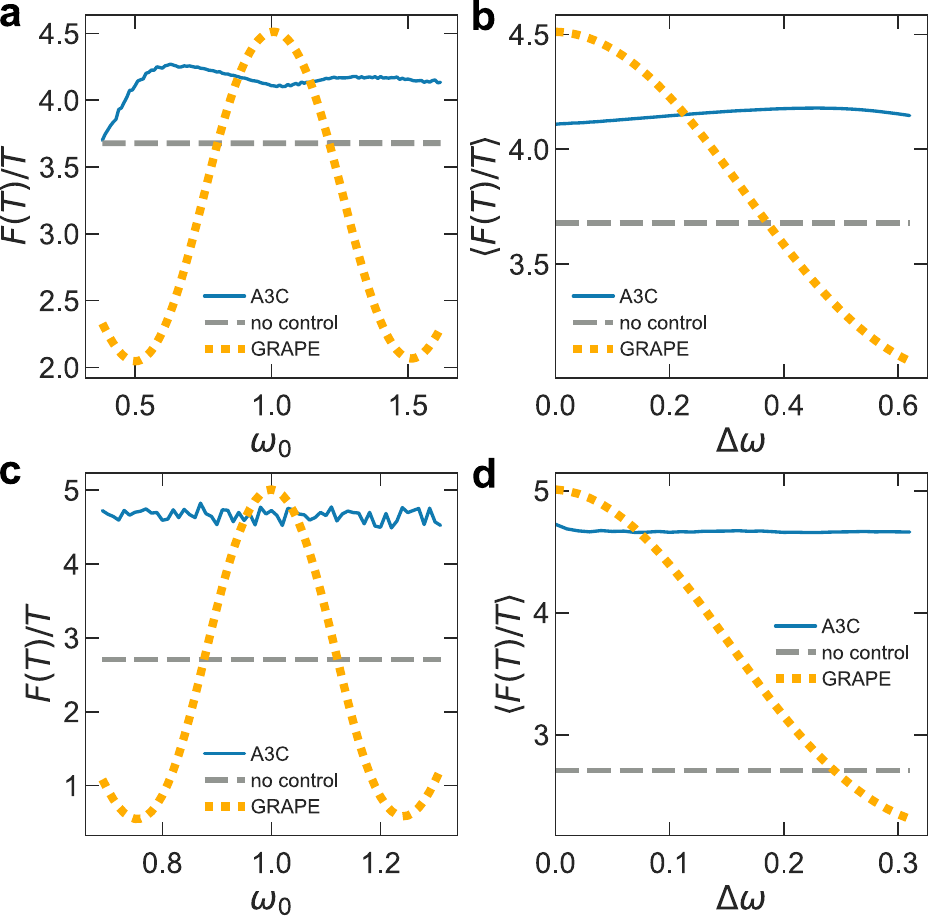}
  \caption{ \text{\black{Generalizability} of the control under spontaneous emission.} \textbf{a}, \textbf{c} $F(T)/T$ v.s. $\omega_0$ for three different methods. Note that the results from the GRAPE method are obtained using the pulses generated for $\omega_0=1$ only, while those from A3C are obtained using a neural network trained at $\omega_0=1$. \textbf{b}, \textbf{d} average $F(T)/T$ in a range $[1-\Delta\omega,1+\Delta\omega]$ corresponding to the results of \textbf{a} and \textbf{c} respectively. \textbf{a}, \textbf{b} $\Delta T=0.1$, $T=10$; \textbf{c}, \textbf{d} $\Delta T=1$, $T=20$.
  \label{fig:main:xfer_se}
  }
\end{figure}

As \black{in the case of dephasing dynamics}, we consider the \black{generalizability} of different methods in a situation involving $\omega_0$ that distributes uniformly in a range. Again, we use GRAPE to obtain optimal control sequences for $\omega_0=1$ and apply that to other values. For A3C, we trained the neural network at $\omega_0=1$; the resulting sequence is then used to obtain an estimate of the true $\omega_0$ value. A new sequence is then generated using the neural network already trained at $\omega_0=1$ with the estimated $\omega_0$. The best-performing results out of 100 A3C outputs are shown as the blue solid lines in Fig.~\ref{fig:main:xfer_se}, while the results from GRAPE are shown as the orange dotted lines. The left column of Fig.~\ref{fig:main:xfer_se} shows  $F(T)/T$ as functions of $\omega_0$ for two $\Delta T$ values. In both cases, the GRAPE method outperforms A3C in a narrow neighborhood around $\omega_0=1$, but its QFI decreases substantially as $\omega_0$ further deviates. On the other hand, A3C exhibits great \black{generalizability}: for $\Delta T=0.1$ the QFI does not decrease until $\omega_0$ is reduced to $\omega_0\lesssim0.6$, while for $\Delta T=1$ the QFI remains approximately the same for the entire range of $\omega_0$ considered.  The average $F(T)/T$ in the range $[1-\Delta\omega,1+\Delta\omega]$ are shown in the right column of Fig.~\ref{fig:main:xfer_se}. In Fig.~\ref{fig:main:xfer_se}{b}, A3C outperforms GRAPE when $\Delta\omega\gtrsim0.22$, while in Fig.~\ref{fig:main:xfer_se}{d}, A3C outperforms GRAPE in an even larger range $\Delta\omega\gtrsim0.07$.

Overall we conclude that in the case of spontaneous emission, the A3C algorithm provides comparable results to GRAPE, although it cannot give higher QFIs. Nevertheless, A3C has much greater \black{generalizability}, as is consistent with \black{the case concerning the dephasing dynamics}.

\subsection{Sequences with Gaussian Pulses}

\begin{figure*}
  \centering
  \includegraphics[width=0.75\linewidth]{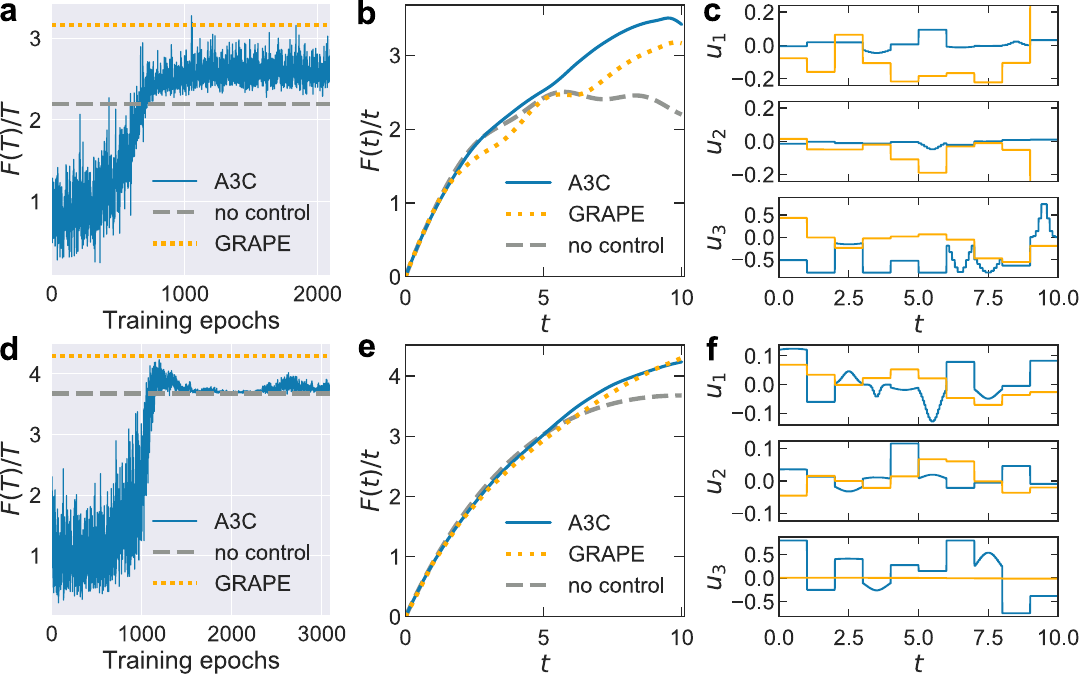}
  \caption{ \text{Quantum parameter estimation  using Gaussian pulses as building blocks for A3C.}  \textbf{a}-\textbf{c} dephasing dynamics with $\vartheta={\mathrm \pi}/4$. \textbf{d}-\textbf{f} spontaneous emission.
  \textbf{a}, \textbf{d} show the learning procedures.
  \textbf{b}, \textbf{e} show $F(t)/t$ for the best training results selected from each case. \textbf{c}, \textbf{f} show the Gaussian pulse profiles, respectively. Note that the GRAPE results shown here use square pulses. Parameters: $\Delta T=1$, $T=10$.
  \label{fig:main:g}
  }
\end{figure*}

For all results shown above, the control sequences involve square pulses only. In practical experiments, shaped pulses are sometimes used. Therefore in this section we consider Gaussian pulses as an example. The total time $T$ is still divided into smaller pieces with $\Delta T$. However, at the $j$th piece the piecewise constant pulse is replaced by a Gaussian centering on that piece and truncated on the ends:
\begin{equation}
u^{(j)}(t)=A^{(j)}\exp\left\{-\left[\left(t-t^{(j)}\right)/\sigma^{{\rm g},(j)}\right]^2\right\},
\end{equation}
where $A^{(j)}$ indicates the amplitude and $\sigma^{{\rm g},(j)}$ the flatness of the pulse. We demonstrate here that with A3C method it is natural to accommodate non-boxcar pulses.

In Fig.~\ref{fig:main:g} we show A3C results using Gaussian pulses and compare them to GRAPE results using square pulses.
Figure~\ref{fig:main:g}{a}-{c} show results under dephasing dynamics with $\vartheta={\mathrm \pi}/4$, and Fig.~\ref{fig:main:g}{d}-{f} results under the spontaneous emission. In both cases $\Delta T=1$, $T=10$. For dephasing dynamics, our best results from A3C outperform GRAPE, as is also the case for square pulses generated by A3C.  For spontaneous emission, our best performing result has a QFI value slightly lower than those from GRAPE with square pulses, but their values are very close. These results indicate that A3C method can naturally accommodate pulses other than square shape. \black{We note that our use of Gaussian pulses is theoretical, and in practical situations, experimentally more relevant ones such as the Blackman pulses \cite{Goerz2014} should be used. These shaped pulses are implemented by introducing constraints to the gradient in GRAPE \cite{Skinner2010} or by modifying the action from the RL agent directly}.

\section{Discussion}
The \black{generalizability} of RL, or sometimes called ``generalization'' in the literature, is an actively studied topic in computer science, for example on problems related to game playing \black{where the RL agent trained under one level of the game can be used to clear other levels} \cite{Pathak2017,Burda2018,Nichol2018,Cobbe2018}. While the reason why RL is \black{generalizable} is not completely clear, one suggestion has it that it likely arises from the underfitting by the neural network to the training data \cite{Mackay2003}, which is supported by studies showing that reducing overfitting improves \black{generalizability} \cite{Cobbe2018}.

The \black{generalizability} in fact has a much wider scope than what has been studied here. In the so-called ``transfer learning'' \cite{Taylor2009},  experiences gained from one training of the RL agent can be used to improve its performance on different but related tasks by, for example, minimal updates of the network parameters. In contrast, our method does not alter network parameters \black{while only generalizes the neural network in new RL environments with different parameters to estimate}. We therefore believe that RL can be made even more \black{generalizable} by further studies involving more sophisticated algorithms. 

To summarize, RL, in particular the A3C algorithm, is capable of finding the control protocol that enhances QFI in a way comparable to the traditionally-used GRAPE method, and is in certain situations superior than GRAPE, e.g. for pulse sequences with larger time steps. Moreover, RL can \black{naturally} accommodate non-boxcar pulse shapes. Nevertheless, the key advantage afforded by RL is the \black{generalizability}, namely the neural network trained for one estimated parameter value can efficiently generate pulse sequences that provide reasonably enhanced QFI for a broad range of parameter values, while in order to achieve the same level of QFI the GRAPE algorithm has to be applied in full each time with a new parameter estimation. 
Our results therefore suggest that RL-based methods can be powerful alternatives to commonly used gradient-based ones, capable to find control protocols that could be more efficient in practical quantum parameter estimation.

\section*{Methods}
In this section we describe the RL framework shown in Fig.~\ref{fig:RL}. We also provide an expansive review of the RL methods and the detail on implementation in the Supplementary Methods.

Figure~\ref{fig:RL}{a} shows the RL agent who takes an action as prescribed by a neural network. In our problem, the action is essentially the control field which steers the qubit according to the master equation, Eq.~\eqref{eq:meq},
and the resulting state of the evolution determines the reward the agent receives. In practice, the reward encodes the QFI, i.e. higher reward will be obtained when greater QFI is given by the control. 

The action taken by the agent implies a time evolution of the quantum state according to Eq.~\eqref{eq:meq} with the control field, $u_k(t)$. All possible actions therefore form a continuous set. We solve this problem using the Actor-Critic algorithm \cite{sutton2018}, as shown in Fig.~\ref{fig:RL}{b}. Such algorithm is particularly suitable to our problem as it can treat continuous actions. The key of the algorithm is that the neural network is not only updated using the reward, but also a state value, the latter of which greatly improves the efficiency of the training procedure.
At certain time step, the neural network takes \black{the density matrix of} the quantum state as an input, and outputs both an action, and a state value which assesses how likely the state will lead to a larger QFI. \black{The state is then evolved using the output action, obtaining the new state and QFI, which is then implemented into the reward. The reward and state value combines into a so-called ``loss function'' that provides feedback, by updating the neural network, for the RL agent to make better decisions. The RL agent takes the new quantum state to repeat the above step until time $T$ is reached, concluding one ``episode'' of training. After that, the quantum state is reset for the next episode to begin with. A completed episode outputs a pulse profile by sequencing the actions taken in each time step}.

In order to improve the efficiency of computation,
we used a parallel version of the Actor-Critic algorithm called Asynchronous Advantage Actor-Critic (A3C) algorithm \cite{Mnih2016}. \black{In this case, several copies of the agent and environment (called local agents and environments) run in parallel, and as each of them finishes one episode, the solution is delivered to a global agent for further optimization. The optimal policy among these results is then regarded as the output from one ``epoch'' of training, i.e. one epoch involves several episodes of training from different local agents. Since different local agents deliver their results at different times, the procedure is asynchronous}.
The details of both the Actor-Critic and the A3C algorithm are described in the Supplementary Methods,
as well as the pseudo-code describing the implementation of the algorithm.


\section*{Data availability}
The datasets generated during this study are available from the corresponding author upon reasonable request.

\section*{Code availability}
The code used to generate data is available from the corresponding author upon reasonable request.

\vspace{12pt}

\section*{Acknowledgements}
This work is supported by the Research Grants Council of the Hong Kong Special Administrative Region,
China (Grant Nos.~CityU 21300116, CityU 11303617, CityU 11304018, CUHK 14207717),
the National Natural Science Foundation of China (Grant Nos.~11874312, 11604277, 11874292, 11729402, 11574238),
the Guangdong Innovative and Entrepreneurial Research Team Program (Grant No.~2016ZT06D348), and the Key R\&D Program of Guangdong province (Grant No. 2018B030326001).

\section*{Author contributions}
X.W. and H.Y. conceived the project, H.X. and J.L. performed calculations. All authors discussed the results and implications at all stages and wrote the paper.




\newpage
\onecolumngrid
\vspace{3cm}


\setcounter{secnumdepth}{6}
\setcounter{section}{0}
\setcounter{equation}{0}
\setcounter{figure}{0}
\setcounter{table}{0}
\renewcommand{\theequation}{S-\arabic{equation}}
\renewcommand{\thesection}{S-\Roman{section}}
\renewcommand{\thesubsection}{\Roman{subsection}}
\renewcommand{\thesubsubsection}{\Roman{subsection}--S\arabic{subsubsection}}
\renewcommand{\thefigure}{S\arabic{figure}}
\renewcommand{\thetable}{S-\Roman{table}}

\renewcommand\figurename{Supplementary Figure}
\renewcommand\tablename{Supplementary Table}
\newcommand\Scite[1]{[S\citealp{#1}]}
\newcommand\Scitetwo[2]{[S\citealp{#1}, S\citealp{#2}]}
\newcommand\citeScite[2]{[\citealp{#1}, S\citealp{#2}]}
\newcolumntype{M}[1]{>{\centering\arraybackslash}m{#1}}
\newcolumntype{N}{@{}m{0pt}@{}}
\makeatletter \renewcommand\@biblabel[1]{[S#1]} \makeatother

\subsection{Supplementary Methods}
\subsubsection{Reinforcement learning}
\label{sec:rl_basic}

The  Reinforcement Learning (RL) framework is schematically shown in Fig.~\ref{fig:RL}{a}. The key ingredients of the RL process include a state space $\mathcal{S}$, an action space $\mathcal{A}$, and a reward $\mathcal{R}$ \cite{sutton2018}. In the RL procedure, an agent at state $s_j\in\mathcal{S}$ chooses an action $a_j\in\mathcal{A}$ according to a probabilistic policy $\pi_\theta(a_j|s_j)$ where $\theta$ represents parameters of the policy. For example, when using a neural network to represent the policy, $\theta$ represents the weights and biases of the neural network. The action $a_j$ results in a new state $s_{j+1}$ according to which the agent receives a numerical reward $r_{j+1}\in\mathcal{R}$. For a given optimization problem, one encapsulates the goal of the problem into the calculation of the rewards, as well as relevant constraints in the available states and actions. In practice, the reward for a given state is not only related to its immediate next step, but several steps in its future, so the total discounted reward for $s_j$, a key quantity, is given by
\begin{equation}
    R_j=\sum_{k=1}^{\infty}\alpha^{k-1} r_{j+k},
\end{equation}
where $\alpha\in(0,1]$ is the reward decaying rate indicating the relative weight between adjacent steps in calculating the total discounted reward received at a given step. When $\alpha=1$ the rewards from all future steps contribute equally, while when $\alpha\rightarrow0$ only the immediate next step provides the major contribution. Then, the probability that the agent takes certain action is enhanced or suppressed, according to the value of the total discounted reward. After sufficient iterations of training, the agent learns the optimal actions to take in order to maximize the total discounted reward, thereby gives an optimal solution to the desired problem.

In the RL procedure, the exploration of the agent in the state and action spaces is summarized into a sequence $s_0,a_0,r_1,s_1,a_1,r_2,\ldots,s_k,a_k,r_{k+1},\ldots$, called a \textit{trajectory}. To figure out what is the best action to take at state $s$, we define the state-action value function,
\begin{equation}
    Q^{\pi}(s,a)={\mathbb{E}[R_j|s_j=s, a_j=a]},
    \label{eq:s-a-v-func}
\end{equation}
where the expectation includes discounted rewards of all the trajectories after taking the action $a$ at the state $s$ in the $j$th step of the trajectory, provided that the policy $\pi$ is observed thereafter \cite{sutton2018}. We also define the value of a state to evaluate the likelihood that a given state would lead to a higher reward,
\begin{equation}
    V^{\pi}(s)={\mathbb{E}[R_j|s_j=s]},
    \label{eq:s-v-func}
\end{equation}
where the expectation includes discounted rewards of all the trajectories starting from the state $s$ in the $j$th step, provided that the policy $\pi$ is followed thereafter \cite{sutton2018}.

An RL policy $\pi$ is declared ``optimal'' when the actions selected by the policy in each state are such that the resulting expectation value of discounted rewards for all states $s\in\mathcal{S}$ is no less than that from any other policy $\pi'$, i.e. $V^{\pi}(s)\ge V^{\pi'}(s)$ \cite{sutton2018}. Corresponding to the optimal policy $\pi_{\theta^*}(a|s)$, the optimal value functions are
\begin{eqnarray}
    Q^*(s,a;\theta_{\rm v}^*) &=& \max_{\pi}Q^{\pi}(s,a;\theta_{\rm v}),\\
    V^*(s;\theta_{\rm v}^*) &=& \max_{\pi}V^{\pi}(s;\theta_{\rm v}),
\end{eqnarray}
where the notations $\theta^*$ and $\theta_{\rm v}^*$ represent optimal choices of the neural network parameters for the policy and value functions. If the optimal value functions are known, the RL agent simply chooses the action $a_j$ that has the largest state-action value $Q^{*}(s_j,a_j)$ in state $s_j$. Alternatively, at  state $s_j$ one may choose the next state $s_{j+1}$ that has the largest state value $V^{*}(s_{j+1})$. Thus, there are two ways for an RL algorithm to solve an optimization problem: the agent either learns the optimal policy, or if the policy is otherwise specified, the optimal value functions \Scite{Konda2003}. The two methods are discussed below.

In the so-called value-based method, the RL agent learns optimal value functions. The state value function and state-action value function are solved iteratively using the Bellmann equations,
\begin{eqnarray}
    Q^{*}(s,a;\theta_{\rm v}^*) &=& r+\alpha \max_{a'}Q^{*}(s',a';\theta_{\rm v}^*),\\
    \label{eq:bellmann_q}
    V^{*}(s;\theta_{\rm v}^*) &=& r+\alpha V^{*}(s';\theta_{\rm v}^*),
    \label{eq:bellmann_v}
\end{eqnarray}
where $a'$ represents all possible actions in the next state $s'$ \cite{sutton2018}. We define the loss functions as
\begin{eqnarray}
    L_Q = \left[R_j^n + \alpha^n\max_{a'}Q^{\pi}(s_{j+n},a) - Q^{\pi}(s_j,a)\right]^2, \label{eq:loss_q}\\
    L_V = \left[R_j^n + \alpha^nV^{\pi}(s_{j+n}) -V^{\pi}(s_j)\right]^2,\qquad
    \label{eq:loss_v}
\end{eqnarray}
where $R_j^n=\sum_{k=1}^{n}\alpha^{k-1} r_{j+k}$ is called the ``$n$-step'' return \cite{sutton2018}.
We take the $\varepsilon$-greedy policy commonly used in deep Q-learning network \cite{Mnih2015,Zhang2018} as an example. Under this policy, the RL agent does either of the two things at state $s_j$: with probability $1-\varepsilon$ the agent takes the action $a_j$ that maximizes $Q^{\pi}(s_j,a_j)$, or with probability $\varepsilon\in(0,1]$ an action is randomly chosen. The latter mechanism encourages the agent to explore a wider range in the search space to reach a globally optimal solution. In practice, $Q^{\pi}(s, a)$ in the loss function Eq.~\eqref{eq:loss_q} is the prediction by the neural network and $r+\alpha \max_{a'}Q^{\pi}(s',a')$ is calculated from the trajectories of the RL agent. The training procedure of the neural network is essentially minimization of the loss function, during which the state-action values given by the neural network are improved.

We note that in the value-based algorithm, the policy is fixed, and only the value functions are updated, which may not be sufficient to find a globally optimal solution \Scite{Konda2003}. More importantly, the way of storing the action space and trajectories have assumed that the actions are discrete, and it becomes far more complicated to treat problems with continuous actions, as is the case of control fields. As shall be discussed below, the policy-based algorithm is most suitable for our problem.

The policy-based algorithm directly updates the policy parameters $\theta$ without the need of storing a large amount of RL trajectories. A typical form of the loss function is defined as \Scite{Williams1992}
\begin{equation}
    L= - \sum_j{\log\left(\pi_\theta(a_j|s_j)\right) A_j},
    \label{eq:policy}
\end{equation}
where $A_j$ is the advantage function,
\begin{equation}
    A_j=R_j-b(s_j),
    \label{eq:adv}
\end{equation}
which evaluates the advantage of the chosen trajectory, with the  \textit{baseline} function $b(s_j)$, normally being the estimated state value function, that reduces the variance and speeds up the learning process \cite{sutton2018,Mnih2016}.
When the value of $A_j$ is large for an action $a_j$, minimizing $L$ increases $\pi_\theta(a_j|s_j)$, implying that the probability to choose the action $a_j$ in state $s_j$ is increased.

In our problem, a quantum state is completely described by the density matrix $\hat\rho^{(j)}$ for each time step $j$. Therefore our state in the RL procedure is defined using elements of the density matrix as
\begin{equation}
    \begin{split}
    s_j=&\left(\mathrm{Re}(\hat\rho_{00}),\mathrm{Im}(\hat\rho_{00}),
    \mathrm{Re}(\hat\rho_{10}),\mathrm{Im}(\hat\rho_{10}),\right.\\
    &\left.\mathrm{Re}(\hat\rho_{01}),\mathrm{Im}(\hat\rho_{01}),
    \mathrm{Re}(\hat\rho_{11}),\mathrm{Im}(\hat\rho_{11})\right).
    \end{split}
\end{equation}
Our action space is formed by a set of control fields
$\left(u_1^{(j)},u_2^{(j)},...,u_p^{(j)}\right)\equiv a_j\in\mathcal{A}$, which steers our quantum state $s_j$ to $s_{j+1}$ according to the master equation Eq.~\eqref{eq:meq}.
Evaluation of the new state $s_{j+1}$ and the agent obtains the single step reward:
\begin{equation}\label{eq:reward}
    r_{j+1}=\left\lbrace\begin{array}{ll}
        \frac{F(j+1)-\eta F_0(j+1)}{F_0(j+1)}, &j+1<N,\\
        \frac{F(j+1)-\eta F_0(j+1)}{F_0(j+1)}\times C, &j+1=N,
    \end{array}
    \right.
\end{equation}
where $F$ and $F_0$ are the corresponding QFI from Eq.~\eqref{eq:qfi} with and without control, respectively. $\eta \ge1$ and $C\ge1$ are constant parameters used in the training process. $\eta $ ensures a non-zero reward to the agent in case the RL agent would apply $u_{1,2,3}(t)=0$, while $C$ gives an extra significance to the last evolution step. After an episode of training, the action sequence in each trajectory constitutes the control field. We also note that our choice of the reward function is not unique.

\begin{algorithm}[thb]\label{alg:A3C-PPO}
  \caption{(episodic) Asynchronous advantage actor-critic with clipped surrogate function}
  \begin{algorithmic}[width=0.8\linewidth]
  \State Initialize the global counter $N^{\mathrm{ep}}=0$
  \Repeat{}
      \State Clear gradients: $d\theta\leftarrow0,d\theta_{\rm v}\leftarrow0$
      \State Synchronize thread-specific parameters: $\theta'=\theta$ and $\theta'_{\rm v}=\theta_{\rm v}$
      \State Reset environment and initial state $s_0$
      \Repeat{}
          \State Choose action $a_j$ according to policy $\pi_{\theta'}(a_j|s_j)$
          \State Update state $s_j\leftarrow s_{j+1}$ and receive reward $r_{j+1}$
          \State $j\leftarrow j+1$
      \Until terminal state $s_N,j=N$
      \State $N^{\mathrm{ep}}\leftarrow N^{\mathrm{ep}}+1$
      \State Initialize the thread-specific counter $N^{\mathrm{ppo}}=0$
      \Repeat{}
          \State $R=0$
          \For{$j\in\{N-1,...,0\}$}
              \State $R\leftarrow r_{j+1}+\alpha R$
              \State Accumulate gradients w.r.t. $\theta$:
              \begin{equation}\label{eq:ppo_1}
                  d\theta\leftarrow d\theta
                  + \partial\min\left(
                      \nu_j(\theta)A_j, \mathrm{clip}(\nu_j(\theta),1-\epsilon,1+\epsilon)A_j
                      \right)
                  /\partial\theta
              \end{equation}
              \State Accumulate gradients w.r.t. $\theta_{\rm v}$:
              \begin{equation}\label{eq:ppo_2}
                  d\theta_{\rm v}\leftarrow d\theta_{\rm v}+\partial A_j^2/\partial\theta_{\rm v}
              \end{equation}
      \EndFor
      \State $N^{\mathrm{ppo}}\leftarrow N^{\mathrm{ppo}}+1$
      \State Perform asynchronous update of $\theta$ using $d\theta$ and of $\theta_{\rm v}$ using $d\theta_{\rm v}$.
      \Until{$N^{\mathrm{ppo}}>N^{\mathrm{ppo}}_{\mathrm{max}}$}
  \Until{$N^{\mathrm{ep}}>N^{\mathrm{ep}}_{\mathrm{max}}$}
  \end{algorithmic}
\end{algorithm}

\begin{table*}[tbh]
    \begin{center}
    \begin{tabular}{l l | l l}
        Hyper-parameter (A3C) & Value & Hyper-parameter (``A3C+PPO") & Value \\
        \hline
        RMSProp Learning rate & $10^{-5}$ & Adam Learning rate & $2\times10^{-4}$ \\
        Reward decay factor $(\alpha)$ & 0.99 & Reward decay factor $(\alpha)$ & 0.9 \\
        Entropy weight $(\eta)$ &$10^{-4}$ & Entropy weight $(\eta)$ &$10^{-3}$ \\
        Batch size & $N$, $T/\Delta T$ & Batch size & $N$, $T/\Delta T$\\
        $C$, in reward function & 10 & $C$, in reward function & 10\\
        $\eta$, in reward function & 1.001 & $\eta$, in reward function & 1.001\\
        Maximum gradient norm & 40 & Maximum gradient norm & 40\\
        Maximum amplitudes $(|u_k|_{\mathrm{max}})$ & 4 &Maximum amplitudes $(|u_k|_{\mathrm{max}})$ & 4\\
         &  & PPO clipping $\epsilon$ & 0.12\\
         &  & Num. PPO steps, $N^{\mathrm{ppo}}_{\mathrm{max}}$ & 10\\
    \end{tabular}\par
\end{center}
\caption{The hyper-parameters for A3C and A3C with PPO strategy}
\label{tab:A3C}
\end{table*}

\begin{figure}[t]
    \includegraphics[width=0.5\linewidth]{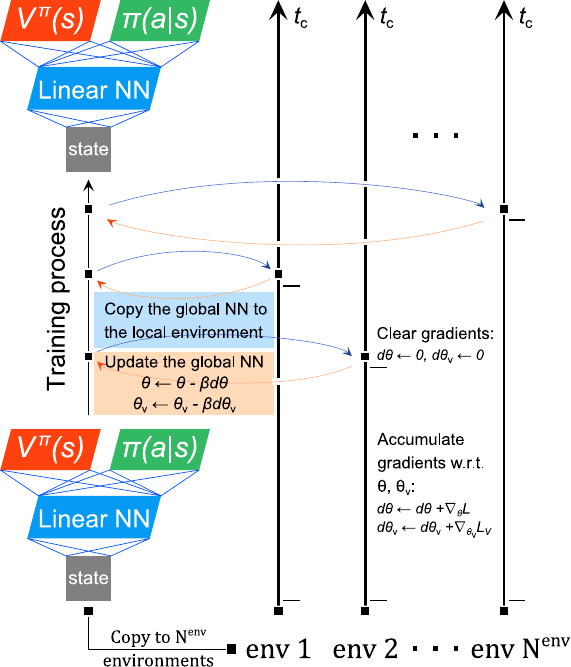}
    \caption{
        Schematics of the A3C algorithm, adapted from \Scite{Seita2018}. The RL neural network is trained asynchronously based on the trajectories of local networks in $N^{\mathrm{env}}$ RL environments, labeled as ``env $i$''. The notation $t_{\rm c}$ is the wall-clock time, $\beta$ is the learning rate. The black dots on the time direction mark the end of each training episode.
    }
    \label{fig:supp:A3C}
\end{figure}

\subsubsection{Actor-critic algorithm}\label{sec:ac_detail}

The Actor-Critic algorithm combines the advantages of policy-based and value-based methods. Figure~\ref{fig:RL}{b} illustrates the basic procedure of the Actor-Critic algorithm. Two neural networks are involved: the actor network governing the policy that chooses actions, and the critic network managing the value functions, which in turn changes the baseline function used in further policy-making \cite{sutton2018}.  More specifically, the state value $V^{\pi}(s)$ generated by the critic network is plugged into Eq.~\eqref{eq:adv},
\begin{equation}
    A_j = R_j^n + \alpha^nV^{\pi}(s_{j+n}) -V^{\pi}(s_j).
    \label{eq:adv_2}
\end{equation}
Note that the ``$n$-step'' return is used instead of $R_j$ so that only the $n$ future steps are involved. This is the key distinction from the policy-based method \cite{sutton2018}.
In the training process, the actor and the critic networks minimize the loss function simultaneously.
We update the critic network through Eq.~\eqref{eq:loss_v} while the actor network is trained through Eq.~\eqref{eq:policy} using the advantage function defined by Eq.~\eqref{eq:adv_2}.

In order to improve the efficiency of the learning process, a parallellized version of Actor-Critic algorithm called A3C, short for Asynchronous Advantage Actor-Critic \cite{Mnih2016}, is implemented in our calculation.

\subsubsection{Asynchronous advantage actor-critic algorithm}

The key structure of Asynchronous Advantage Actor-Critic (A3C) is sketched in Fig.~\ref{fig:supp:A3C}. The desired policy and value functions are generated by the neural network (left column in Fig.~\ref{fig:supp:A3C}), called the ``global'' network. The neural network is composed of the state value network $V^{\pi}(s)$ (orange color), the policy network $\pi(a|s)$ (green color) and the  fully-connected linear layers (blue color). At the beginning of the training process, we made $N^{\mathrm{env}}$ copies of the global network, called  ``local'' networks. Then, each of the local networks is allowed to run in independent RL environments, in which the RL agents, called the ``local" agents, optimize policies and value functions via gradients with respect to the loss functions.
At the end of a training episode for each parallel RL procedure, the local agent uploads the accumulated gradient to update the global network. Then, the updated global network is downloaded back to the local environment, starting a new episode with the environment properly reset. Note that in the entire process, all local agents act independently, which is why the algorithm is asynchronous \citeScite{Mnih2016}{a2c}.

We now give details of our implementation of the A3C algorithm. The RL states are first fed through 4 hidden layers, each composing 200 ReLU units \Scite{Paszke2017}. The resulting outputs are then passed to both the value and policy networks. The value network is constructed by one hidden layer with 200 ReLU units and one fully-connected linear layer outputting a real number as the state value. The policy network has one hidden layer with 200 ReLU units and two fully-connected linear layers as output layers. The outputs are six real numbers $\mu_k$, $\sigma^{\rm G}_k$, $k=1,2,3$ forming three normal distributions $N(\mu_k, \sigma^{\rm G}_k)$. Here, $\mu_k$ is modified by the SoftShrink$(\lambda)$ activation function with $\lambda=0.25$ and $\sigma^{\rm G}_k$ is modified by the SoftPlus activation function \Scite{Paszke2017}. The continuous actions $u_k$ are randomly sampled from those normal distributions.

We use the differentiation of the normal distribution as the entropy regularization term, $-\frac{1}{2}(\log(2\pi\sigma^2)+1)$, to encourage the agent to explore the entire search space. We use the RMSProp optimizers with shared parameters that are updated asynchronously among parallel environments \cite{Mnih2016}. We keep the choice of hyper-parameters which are listed in the left column of Table \ref{tab:A3C} similar to those used in \cite{Mnih2016}. The pseudocode for A3C can be found in \cite{Mnih2016}. Next we will discuss an optimized version of the code, i.e. with Proximal Policy Optimization (PPO) algorithm \citeScite{Schulman2017}{Heess2017}.

Generally, optimization with the logarithm of the policy gradient leads to large policy updates which, in some cases, makes the learning process unstable. The Proximal Policy Optimization (PPO) algorithm replaces the logarithm in Eq.~\eqref{eq:policy} with the probability ratio between the old and the new policy:
\begin{equation}
    \nu_j(\theta)=\frac{\pi_{\theta}(a_j|s_j)}{\pi_{\theta_{\mathrm{old}}}(a_j|s_j)},
\end{equation}
and the loss function is also truncated at certain values of  the probability ratio \cite{Schulman2017}. Algorithm \ref{alg:A3C-PPO} shows the pseudocode for the A3C algorithm utilizing the PPO strategy. In this algorithm, we replace the global RMSProp optimizer with the thread-specified Adam optimizers \Scite{Paszke2017}. The right column of Table \ref{tab:A3C} lists the hyper-parameters in the A3C algorithm with PPO strategy.

We have used PyTorch \Scite{Paszke2017} to implement the algorithms and QuTip \Scitetwo{Johansson2012}{Johansson2013} to obtain numerical solutions of Eqs.~\eqref{eq:meq}-\eqref{eq:qfi}.
We also note that practically, when $\Delta T=1$, we have to set smaller learning rates, gradient norm, entropy weight and $N^{\mathrm{ppo}}_{\mathrm{max}}$.

\subsection{Supplementary Discussion}
\subsubsection{Computational complexity}

\begin{figure}[t]
    \includegraphics[width=0.35\linewidth]{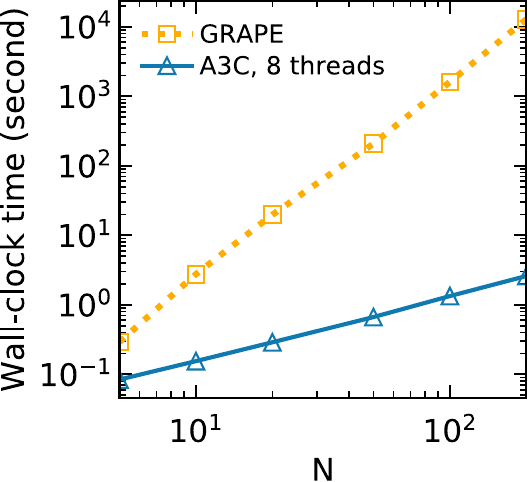}
    \caption{
        Comparison of time complexities between A3C and GRAPE. The x-axis shows the system size, i.e. the number of time steps $N$. The y-axis shows the wall-clock time cost in seconds during one training epoch of A3C or one iteration of GRAPE. Notice that the base-10 log scale is used for both axes. Eight RL agents in parallel threads are employed in A3C. The algorithms are run on a PC with eight Intel Core i7-7700 CPU (3.60GHz) cores.
    }\label{fig:supp:complexity}
\end{figure}

In our discussion, the computational complexity refers to the time complexity which depends on the number of elementary operations performed during the execution of the algorithm. For the optimal control problem we considered, the evolution time between $0$ and $T$ is discretized into $N$ equal time steps. In most cases, we employ piecewise constant pulse sequences so if we want to compute the evolution of a density matrix from time 0 to $T$ in $N$ time steps with piecewise constant pulses, we need to compute the master equation at $N$ time intervals and the time complexity scales with $N$. Accordingly, we compare the time complexity of A3C and GRAPE with respect to a system size of $N$.

In one episode of A3C, we take the probe state as the input to the RL algorithm, which keeps running until time $T$ is reached. During this process, we have used the master equation $N$ times in the RL environment. The computational complexity is therefore $\mathcal{O}(N)$. On the other hand, the time cost of training the neural network is dependent on the network structure (number of neurons, layers etc.) which is irrelevant to GRAPE. Therefore for the purpose of comparing to GRAPE, the time complexity $\mathcal{O}(N)$  includes the cost of training which adds a prefactor dependent on the details of the network. For GRAPE, according to the analytical results of the gradient of QFI in Ref.~\cite{Liu2017}, we need to compute the evolution of the density matrix $N^2$ times to numerically evaluate the gradient with respect to the control $u_k^{(j)}$ at time step $j$. Thus, computing the gradient of QFI with respect to $u_k^{(j)}$ causes the complexity $\mathcal{O}(N^2)$. During one iteration of GRAPE, we want to update $N$ piecewise controls so the complexity further increases to $\mathcal{O}(N^3)$. One should note that optimizing the QFI is computationally more expensive than optimizing the fidelity with GRAPE \cite{Khaneja2005}. 

We verify our results on a PC with the standard multi-core CPU and plot their wall-clock time costs as functions of the system size $N$ in Fig.~\ref{fig:supp:complexity}. It shows that the wall-clock time costs in one training epoch of A3C and one iteration of GRAPE follows the scaling $\mathcal{O}(N)$ and $\mathcal{O}(N^3)$ respectively, as expected.

We now count into the number of training epochs (A3C) or iterations (GRAPE). In actual implementations, the number of training epochs in the A3C algorithm is usually $\sim10^3$, while the number of iterations in GRAPE is typically between $10^1$ and $10^2$. For small $N$ ($N\lesssim 10$), a full execution of GRAPE can be faster than A3C due to its smaller prefactor of the number of iterations. However, this case corresponds to a larger $\Delta t$ for which we know that the result of A3C may outperform GRAPE in QFI. Therefore we summarize the comparison as follows: For small $N$, A3C may be slower than GRAPE but can produce results with higher QFI, and is more generalizable. For large $N$, A3C is overall faster than GRAPE, producing results with QFI comparable (but not exceeding) GRAPE, and is more generalizable. We believe it is fair to conclude that A3C is more efficient in more experimentally relevant cases, i.e. having larger $N$ or when generalizability is desired.

\subsubsection{Additional results on dephasing dynamics}
\label{sec:more_train}

\begin{figure*}[tbh]
    \begin{center}
    \includegraphics[width=0.9\linewidth]{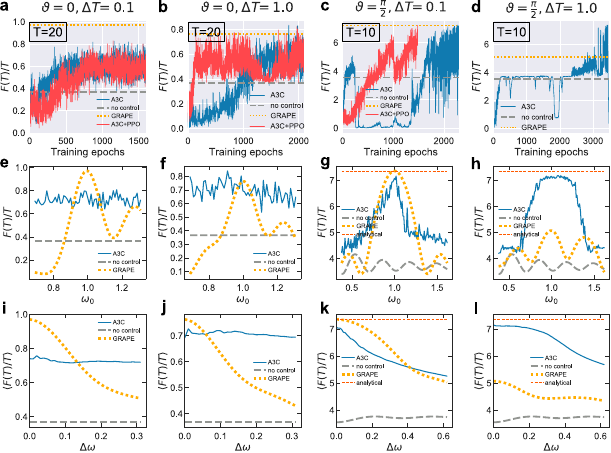}
    \end{center}
    \caption{Quantum parameter estimation under dephasing dynamics along two different axes using square pulses. First column: parallel dephasing $(\theta=0)$ and $\Delta T=0.1$. Second column: parallel dephasing with $\Delta T=1$. Third column: transverse dephasing $(\theta=\pi/2)$ and $\Delta T=0.1$. Fourth column: transverse dephasing with $\Delta T=1$.  The upper row shows the learning procedure, i.e. $F(T)/T$ as functions of training epochs. The middle row compares  $F(T)/T$ v.s. $\omega_0$ for different methods. The bottom row shows average $F(T)/T$ in a range $[1-\Delta\omega,1+\Delta\omega]$. The total times $T$ are indicated.
    }
    \label{fig:supp:train}
\end{figure*}

In the main text, we have provided results of quantum parameter estimation under dephasing dynamics along a chosen axis in Fig.~\ref{fig:main:dp}, i.e.~$\vartheta=\pi/4$. Here, we present results along two other axes: parallel depasing $(\vartheta=0)$ and transverse dephasing $(\vartheta=\pi/2)$. In Fig.~\ref{fig:supp:train}, the training process is shown in the upper row, $F(T)/T$ v.s.~$\omega_0$ the middle row and the average $F(T)/T$ in $[1-\Delta\omega,1+\Delta\omega]$ in the bottom row.  For parallel dephasing, our results are very similar to $\vartheta=\pi/4$ results shown in the main text, namely $F(T)/T$ calculated from A3C is lower than that from GRAPE only in a narrow range of $\Delta\omega$. For $\Delta T=0.1$, A3C outperforms GRAPE when $\Delta\omega\gtrsim0.15$, while for $\Delta T=1$, A3C is better than GRAPE in a wider range, $\Delta\omega\gtrsim0.05$. For transverse dephasing, the situation is slightly more complicated (note that  analytical solutions \cite{Liu2017} are provided as references). When $\Delta T=1$, results from GRAPE has very low $F(T)/T$, thus A3C always outperforms GRAPE. However, for $\Delta T=0.1$, A3C does not possess considerable advantages. For $0\le\Delta\omega\lesssim0.4$, the A3C results have lower $F(T)/T$ than GRAPE, albeit being very close. For $\Delta\omega\gtrsim0.4$, the A3C results is only slightly higher than GRAPE. These calculations therefore suggest that the generalizability of our method is superior as compared to GRAPE in most situations, in particular for cases with larger time step $(\Delta T)$. Nevertheless, in some situations, usually associated with smaller $\Delta T$, our method would not provide considerable improvement. One therefore has to be judicious in choosing appropriate methods for a specific problem. For example, if generalizability is not desired, GRAPE may be more appropriate for pulse sequences with smaller time steps. On the other hand, if pulse sequences have larger time steps, or generalizability becomes important in the problem, the A3C method is desired.


\begin{thebibliography}{1}
\bibitem{a2c}
  \bibinfo{note}{We note that ``asynchronism" is not a necessary condition since
    one may train the agents synchronously using batched experiments in a
    parallel fashion {\Scite{Wu2017}}}.
\end{thebibliography}
\end{document}